\documentclass[onecolumn,english,preprint,floatfix]{revtex4}
\usepackage[T1]{fontenc}
\usepackage[latin9]{inputenc}
\usepackage{amssymb}
\usepackage{esint}

\usepackage{babel}

\usepackage{graphicx}
\usepackage{hyperref}
\usepackage{pdfsync}

\usepackage{amssymb}

\def\tr{\mbox{Tr}\,}

\newcommand{\ee}{\end{equation}}

\newcommand{\be}{\begin{equation}}

\newcommand{\bea}{\begin{eqnarray}}

\newcommand{\eea}{\end{eqnarray}}
\newcommand{\eu}{{\rm e}}

\newcommand{\de}{{\displaystyle\rm\mathstrut d}}

\def\XXint#1#2#3{{\setbox0=\hbox{$#1{#2#3}{\int}$}
     \vcenter{\hbox{$#2#3$}}\kern-.5\wd0}}

\newcommand{\ba}{\begin{eqnarray}}
\newcommand{\ea}{\end{eqnarray}}

\begin{document}

\title{On the Spontaneous Breaking of $U(N)$ symmetry in invariant Matrix Models}

\author{Fabio Franchini}
\affiliation{INFN, Sezione di Firenze, Via G. Sansone 1, 50019 Sesto Fiorentino (FI), Italy}
\affiliation{SISSA and I.N.F.N,  Sezione di Trieste, Via Bonomea 265, 34136, Trieste, Italy}
\affiliation{Department of Physics, Massachusetts Institute of Technology, Cambridge, MA 02139, USA}

\preprint{MIT-CTP/4576}

\date{\today}

\begin{abstract}
Matrix Models are the most effective way to describe strongly interacting systems with many degrees of freedom. They have proven successful in describing very different settings, from nuclei spectra to conduction in mesoscopic systems, from holographic models to various aspects of mathematical physics.
This success reflects the existence of a large universality class for all these systems, signaled by the Wigner-Dyson statistics for the matrix eigenvalues.
These models are defined in a base invariant way and this rotational symmetry has traditionally been read to imply that they describe extended system.
In this work we show that certain matrix models, which show deviations from the Wigner-Dyson distribution, can spontaneously break their rotational ($U(N)$) invariance and localize their eigenvectors on a portion of the Hilbert space. This conclusion establishes once more a direct connection between the eigenvalue and eigenvector distributions.
Recognizing this loss of ergodicity discloses the power of non-perturbative techniques available for matrix models to the study of localization problems and introduces a novel spontaneous symmetry breaking mechanism. 
Moreover, it brings forth the overlooked role of eigenvectors in the study of matrix models and allows for the consideration of new types of observable.


\end{abstract}

\maketitle

Much of our understanding of Physics in the last century has been built on the progressive appreciation of the role of symmetries, from the classification of different phases of matter, to the description of the fundamental (and effective) forces of nature. A special role in this journey has been played by the realization that a system can spontaneously break the symmetry that characterizes it.
The prototypical example of this phenomenon is ferromagnetism: although the system is invariant under rotations, at sufficiently low temperatures the majority of the microscopic magnetic moments align along a particular direction. In gauge theories, the spontaneous breaking of a symmetry (SSB) is accompanied by the by Higgs mechanism, which gives masses to the mediator of the broken symmetries.

SSB mechanisms have been studied in virtually all branches of physics, but, so far, not in a matrix model formulation. Matrix models are well suited to describe systems with many interacting degrees of freedom. They were originally introduced by Wigner in the fifties to describe the spectra of heavy nuclei, under the hypothesis that the Hamiltonian governing the interactions could be well approximated by a randomly generated matrix \cite{wigner51}. The success of this hypothesis has led over the years to the application of matrix models to virtually every branch of physics and beyond: from nuclear physics to string theory, from 2-D quantum gravity to condensed matter physics, from statistical physics to econo-physics, from neuroscience to chaos theory, from number theory to integrable systems, and so on \cite{mehtabook, JPAbook, natobook, akemannbook}. In the different contexts the matrices can correspond to very different physical quantities: in some cases they are taken as dynamical fields with many internal degrees of freedom; in some other instances they are interpreted as links between points in a network (or a discretized manifold); in other situations they are matrix representation of the operators of a many-body quantum theory; and so on.

Despite the differences in physical interpretation of these models, they all share the same mathematical structure, which has offered the opportunity for cross-fertilization between different areas of physics and mathematics over the years. The central object defining the model is the probability distribution function (PDF) $ P ( {\bf M} )$, which provides the weight for the $N \times N$ matrix ${\bf M}$. In the following, we will assume ${\bf M}$ to be hermitian, although other symmetry classes can be considered as well. 
In most applications one assumes basis invariance, so that the PDF can be written as
\be
   P ( {\bf M} ) \propto \eu^{- N \tr V ( {\bf M} )} \; ,
   \label{Pof}
\ee 
and is specified by the potential $V(x)$. 
One important reason for working with this type of PDF is that invariant models enjoy powerful mathematical approaches which, over the years, have led to the calculation of a variety of observables and their matching with empirical measurements in the different contexts of application. \cite{mehtabook, akemannbook}. 

A key property of invariant models is the possibility of factorizing the eigenvalues from the rotational degrees of freedom. That is, writing ${\bf M} = {\bf U}^\dagger {\bf \Lambda} {\bf U}$, where $ {\bf \Lambda}$ is diagonal and ${\bf U}$ is the unitary matrix encoding the eigenvectors of ${\bf M}$, the PDF does not depend on ${\bf U}$. This independence has been interpreted to imply that ${\bf U}$ is uniformly distributed over an $N$-dimensional sphere. Since each eigenvector can extend over the whole Hilbert space of the system, invariant matrix models are supposed to describe {\it extended (i.e. conducting) phases}.
For a matrix ${\bf U}$ belonging to the unitary $U(N)$ group, the distribution of its entries is the {\it Porter-Thomas distribution} \cite{porter56}
\be
  {\cal P} \left( \left| U_{i j} \right|^2 \right) =  \chi \exp \left[ - \chi \left| U_{i j} \right|^2 \right] \; ,
  \label{PTd}
\ee
with $\chi =N$, which means that the mean amplitude of each eigenvector entry is $1/N$ and it is equal to its standard deviation.

Very much like for the ferromagnet, we are about to show that in invariant models some rotational degrees of freedom of the eigenvectors can be frozen by the eigenvalue distribution. In particular, a finite gap between two eigenvalue sets prevents the mixing between the corresponding eigenvectors. Interpreting the matrix as a Hamiltonian, for instance, this eigen-energy structure can emerge if a many-body system has a large degeneracy which is broken by some hyperfine coupling, so that the spectrum is composed by bands separated by gaps. The coupling between the eigenvector and eigenvalue distributions represents a novel SSB mechanism, for which the $U(N)$ symmetry is spontaneously broken into the product of smaller unitary groups $U(n)$s. The critical and universal behavior at the threshold when a gap in the eigenvalue density is about to open has been observed in the literature and discussed in terms of the eigenvalue behavior \cite{eynard06, claeys08}, but now our work characterizes it in terms of symmetries being broken. The idea that the $U(N)$ symmetry could be spontaneously broken in matrix models was originally introduced in \cite{canali95} for a class of models with eigenvalue statistics intermediate between Wigner-Dyson and Poissonian \footnote{This idea has also circled the string theory community, following \cite{bonnet00}}. A few years later, in \cite{pato00}, these models were shown to break the symmetry to a $U(1)^N$ in a particular limit, but a general SSB mechanism has never been discussed before now.  This development unlocks the possibility of analyzing new non-trivial properties of invariant matrix models and to apply them to new phenomena, such as partially localized systems. Moreover, matrix models arise naturally in the description of strongly interacting theories, including string theories and holographic models. Thus, the description of their SSB mechanism is a new ingredient that can be applied to the study of new observables.

\section{General Considerations}

The SSB mechanism in matrix models is easily understood by noticing that the (Haar) integration measure for matrices is flat in matrix  elements space, but curved in the eigenvalue/eigenvectors representation, with a curvature induced just by the eigenvalues \cite{mehtabook}. The line element is given by
\be
   \de s^2 = \tr \left( \de M \right)^2  =
    \sum_{j=1}^N \left( \de \lambda_j \right)^2 
    + 2 \sum_{j > l}^N  \left( \lambda_j - \lambda_l \right)^2 \left| \de A_{jl} \right|^2 \; .
    \label{lineel}
\ee
where $\de  {\bf A} \equiv {\bf U}^\dagger \de {\bf U}$.
Keeping the eigenvalues $\lambda_j$ fixed, the angular degree of freedom $\de {\bf A}_{jl}$ lives on a sphere of radius $r_{jl}=\left| \lambda_j - \lambda_l \right|$. If the two eigenvalues are very distant ($r_{jl} \gg 1$), even a small angular change $\de {\bf A}_{jl}$ can move to a very distant point, producing a large $\de s$.

This geometric argument is at the heart of the SSB mechanism, as can be understood by introducing a fictitious dynamics in the model, in the form of the Brownian motion originally proposed by Dyson \cite{dyson62}. In this picture, one introduces a random noise and lets the eigenvalue and eigenvectors evolve to the equilibrium distribution. The stochastic differential system for this Brownian motion is \cite{allez13}
\bea
    \de \lambda_j & = & - {\de V (\lambda_j) \over \de \lambda_j} \: \de t +
    {1 \over N} \sum_{l \ne j} {\de t \over \lambda_j - \lambda_l} 
    + {1 \over \sqrt{N}} \: \de B_j (t) \: ,
   \nonumber  \\
    \de \vec{U}_j (t) & = & 
    - {1 \over 2 N} \sum_{l \ne j} {\de t \over (\lambda_j - \lambda_l)^2} \:  \vec{U}_j 
    + {1 \over \sqrt{N}} \: \sum_{l \ne j} {\de W_{jl}(t) \over \lambda_j - \lambda_l} \:  \vec{U}_l\: , \nonumber \\
    \label{stocvec}
\eea
where $\de B_j$ and $\de W_{jl}=\de W_{lj}^*$ are delta-correlated, independent, stochastic sources, and $\vec{U}_j$ denotes the $j$-th column of ${\bf U}$, that is, the eigenvector of $\lambda_j$.
We notice that, while the eigenvalues dynamics is self-consistent, the motion of the eigenvectors depends on eigenvalue distribution through their distances. This effect is the source of the coupling between the eigenvectors and the eigenvalues which is not apparent in the equilibrium measure. We remind that eigenvalues are spread on two characteristic scales: while neighboring ones are separated by level repulsion at a distance of the order of $1/N$, the whole distribution spans a length of order $1$. Thus, we see that if two sets of eigenvalues are separated by a gap of the order of unity, eq. (\ref{stocvec}) indicates that the evolution of the eigenvectors toward the subspace spanned by eigenvectors belonging to the distant eigenvalues is suppressed and forbidden in the $N \to \infty$ limit. In this case, the eigenvectors cannot spread ergodically over the whole Hilbert space. The freezing of certain degrees of freedom is an indication of the spontaneous breaking of $U(N)$ symmetry.

\section{SSB in a double well model}

Equipped with these general geometric considerations, we can provide a more concrete construction. In doing so, we will consider the specific example of a double well matrix model, that is, a potential in (\ref{Pof}) of the form
\be
   V_{\rm 2W} (x) = {1 \over 4} x^4 - {t \over 2} x^2 \; ,
  \label{V2W}
\ee
which is known to have a disjoint support for the eigenvalues for $t>2$ \cite{bleher03, claeys06}. That is, for sufficiently deep and separated wells, half of the eigenvalues accumulate around one minimum and the other half around the other minimum\footnote{We will take $N$ to be even, in order to avoid the instatonic oscillation of a single eigenvalue between the two wells \cite{bonnet00}.}. We intent to show that the gap separating the negative eigenvalues from the positive ones induces a breaking of the $U(N)$ symmetry into $U(N/2) \times U(N/2)$.

\paragraph*{Symmetry Breaking Term:} One of the most direct ways to detect a SSB is to introduce an explicit symmetry breaking term and to remove it only after the thermodynamic limits. The most natural choice for this term in matrix models would be $\tr \left( \left[ {\bf M} , {\bf S} \right] \right)^2$, with ${\bf S}$ a given Hermitian matrix. Such a contribution favors the alignment of the eigenvectors of ${\bf M}$ along those of ${\bf S}$ (which thus provides a reference basis). Unfortunately, this choice turns out too complicated to handle. 

Thus, we will consider the following perturbation of the invariant PDF
\be
   P ( {\bf M} ) \propto \eu^{- N \left[ \tr V_{\rm 2W} ( {\bf M} ) 
   + J \left|\tr \left( {\bf \Lambda} \: {\bf T} - {\bf M} \: {\bf S} \right) \right|  \right] }\; ,
   \label{PofB}
\ee 
where the symmetry breaking term serves the same purpose of favoring the alignment of the eigenvectors of ${\bf M}$ with those of ${\bf S}$. Here, $J$ acts as a source strength and ${\bf T}$ is a diagonal matrix with the eigenvalues of ${\bf S} = {\bf V}^\dagger {\bf T} {\bf V}$. If the eigenvectors of ${\bf S}$ and ${\bf M}$ coincide (${\bf V} = {\bf U}$), the symmetry breaking term vanishes. Otherwise, it contributes with a finite amount, which the absolute value insures to be positive (and hence penalizing).
In fact, the absolute value in (\ref{PofB}) can be removed by sorting the eigenvalues of ${\bf M}$ and ${\bf S}$ in increasing order, since it can be checked that this choice insures that $\tr \left( {\bf \Lambda} \: {\bf T} - {\bf M} \: {\bf S} \right)$ is never negative.  In order to induce the correct symmetry breaking, it is convenient to take $N/2$ eigenvalues of ${\bf S}$ to be equal to $t$ and the remaining $N/2$ equal to $-t$ (without loose of generality, we can further set $t=1$). This choices preserve $U(N/2)$ rotational freedom for the eigenvectors corresponding to eigenvalues within one well, while it penalizes rotations in the remaining directions\footnote{It is also possible to work with a non-degenerate ${\bf S}$, since the $J \to 0$ limit at the end select the correct symmetry breaking pattern corresponding to the potential in (\ref{Pof}). However, our choice makes the calculation more transparent.}.

We can now consider the generating function
\be
   W(J) = \ln \int \de {\bf M}  \eu^{- N \tr V_{\rm 2W} ( {\bf M} ) 
   + J N \left| \tr \left( {\bf \Lambda} \: {\bf T} - {\bf M} \: {\bf S} \right) \right|  }  \;,
   \label{genfunc}
\ee
whose derivative with the respect to the coupling $J$ provides the expectation value of the order parameter
\be
   \left. {\de W (J) \over \de J} \right|_{J=0} =
   \langle \left| \tr \left( {\bf \Lambda} \: {\bf T} - {\bf M} \: {\bf S} \right) \right| \rangle \; .
   \label{ordpar}
\ee
This order parameter measures the misalignment between the eigenvectors of ${\bf M}$ and ${\bf S}$.
Note that, contrary to the usual convention, with our definitions the breaking of the symmetry is signaled by a vanishing order parameter, while a non-vanishing one indicates a symmetry restoration (thus (\ref{ordpar}) is in fact a {\it disorder parameter}).
We find that, in the $J \to 0$ limit the order parameter remains finite as long as $N$ is finite. However, taking the $J \to 0$ limit after the thermodynamic limit $N \to \infty$ yields a vanishing order parameter, indicating that the $U(N)$ symmetry has been spontaneously broken into $U(N/2) \times U(N/2)$.

\begin{figure}
\begin{center}
\includegraphics[width=\columnwidth]{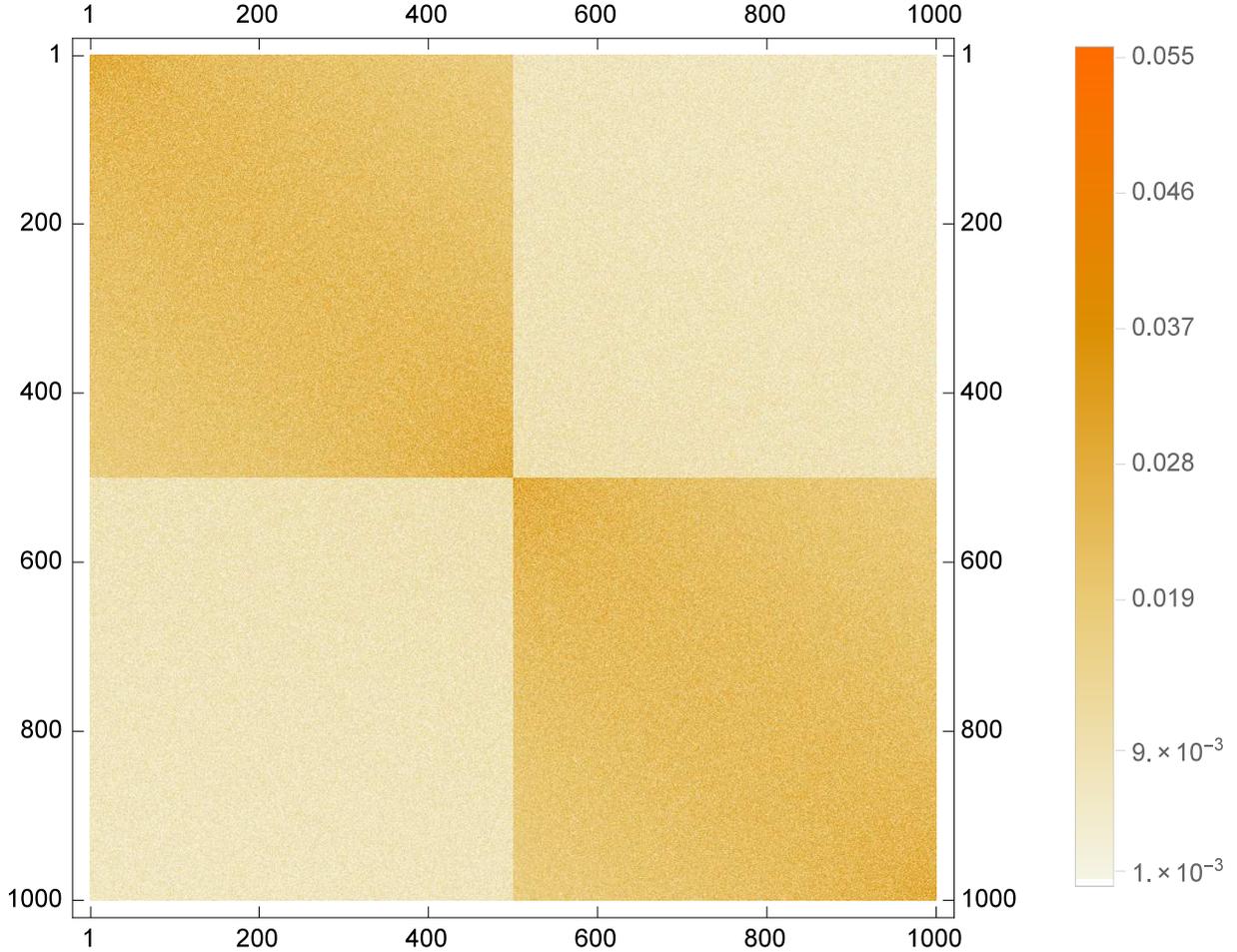}
\caption{Block structure of a spontaneously broken unitary matrix for the double well case: the squared absolute value of the matrix entries clearly show the off-diagonal blocks to be suppressed compared to the diagonal ones. This is a typical results for the unitary matrix $\tilde{\bf U}$ diagonalizing ${\bf M} + {\bf \Delta M}$ in the basis where ${\bf M}$ is diagonal. ${\bf \Delta M}$ has $n \times N$ (with $n=200$ and $N=1000$, being the rank of ${\bf M}$) non-zero elements randomly drawn from a Gaussian distribution, centered around $0$ and variance $N$.}  
\label{fig:2BlocksU}
\end{center}
\end{figure}

Moreover, the calculation shows that, at finite sizes, corrections to the symmetry broken configuration appear in the form $\exp \left[ - N \; J \; \delta \lambda_{j,l} \right]$, where $\delta \lambda_{j,l} \equiv \left| \lambda^{(1)}_j - \lambda^{(2)}_l \right|$ is the distance between two eigenvalues in different wells. These contributions have a natural interpretation in terms of instantons which exchange the two eigenvalues between the wells and progressively restore the broken symmetries. The $J \to 0$ limit renders these instantons  ``massless''  and thus restores the symmetry to the full $U(N)$. In contrast, the $N \to \infty$ limit suppresses these contributions completely and induces the SSB. Note that $\delta \lambda_{j,l}$ is of the order of unity because the eigenvalues belong to different wells. Within a single well, $\delta \lambda_{j,l}$ would be of the order of $1/N$ and thus these instanton contributions would not disappear in the thermodynamic limit and there would be no SSB.

\paragraph*{Finite Size Analysis:} Having established the existence of a SSB mechanism in matrix models, we are now going to provide a numerical procedure to detect it. In the spirit of (\ref{stocvec}), we ask how the eigenvectors respond to a perturbation. In a Gaussian matrix models, they would rotate freely over the $N$-dimensional sphere and the new eigenvectors would quickly acquire a non-vanishing overlap with virtually all the original, non-perturbed, eigenvectors. In a double well model, instead, the SSB confines each set of eigenvectors to rotate in the $N/2$-dimensional sphere spanned by the eigenvectors of each well, with only suppressed overspill in the remaining Hilbert space, vanishing in the thermodynamic limit.
Thus, starting from a matrix ${\bf M}$ representative of the PDF with (\ref{V2W}), we add a perturbation, in the form of a sparse, hermitian Gaussian matrix ${\bf \Delta M}$ with order of $N$ non-zero elements, sampled from an independent Gaussian distribution of width $\sqrt{N}$. Finally we look at the unitary matrix $\tilde{\bf U}$ that connects the eigenvectors of ${\bf M} + {\bf \Delta M}$ with those of the unperturbed ${\bf M}$. 

\begin{figure}
\begin{center}
\includegraphics[width=\columnwidth]{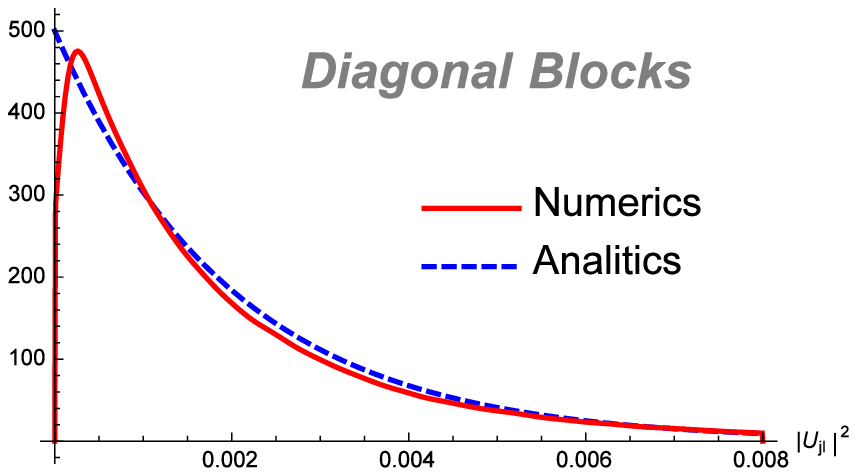}
\includegraphics[width=\columnwidth]{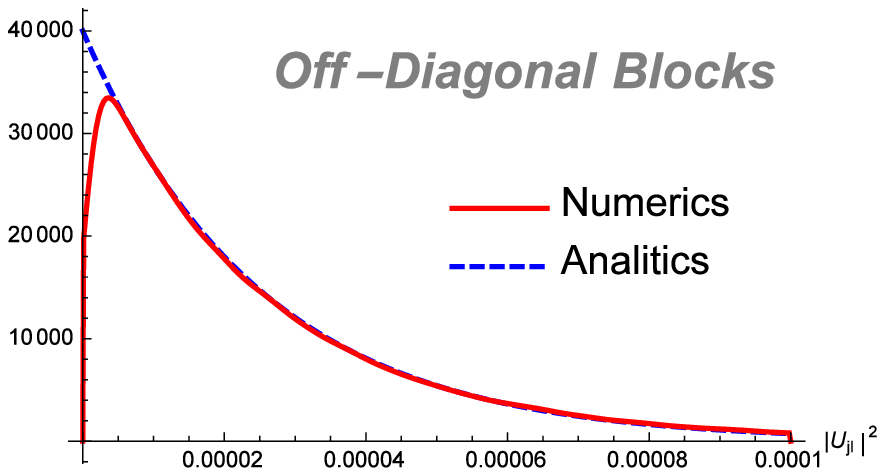}
\caption{Distribution of the squared absolute value entries of the diagonal (Top) and off-diagonal (Bottom) blocks of the matrix in Fig. \ref{fig:2BlocksU}. In red the numerics, while the blue lines are the Porther-Thomas distribution (\ref{PTd}), but with the effective rank $\chi$ given by (\ref{mPTd}).}  
\label{fig:distrib}
\end{center}
\end{figure}

\begin{figure}
\begin{center}
\includegraphics[width=\columnwidth]{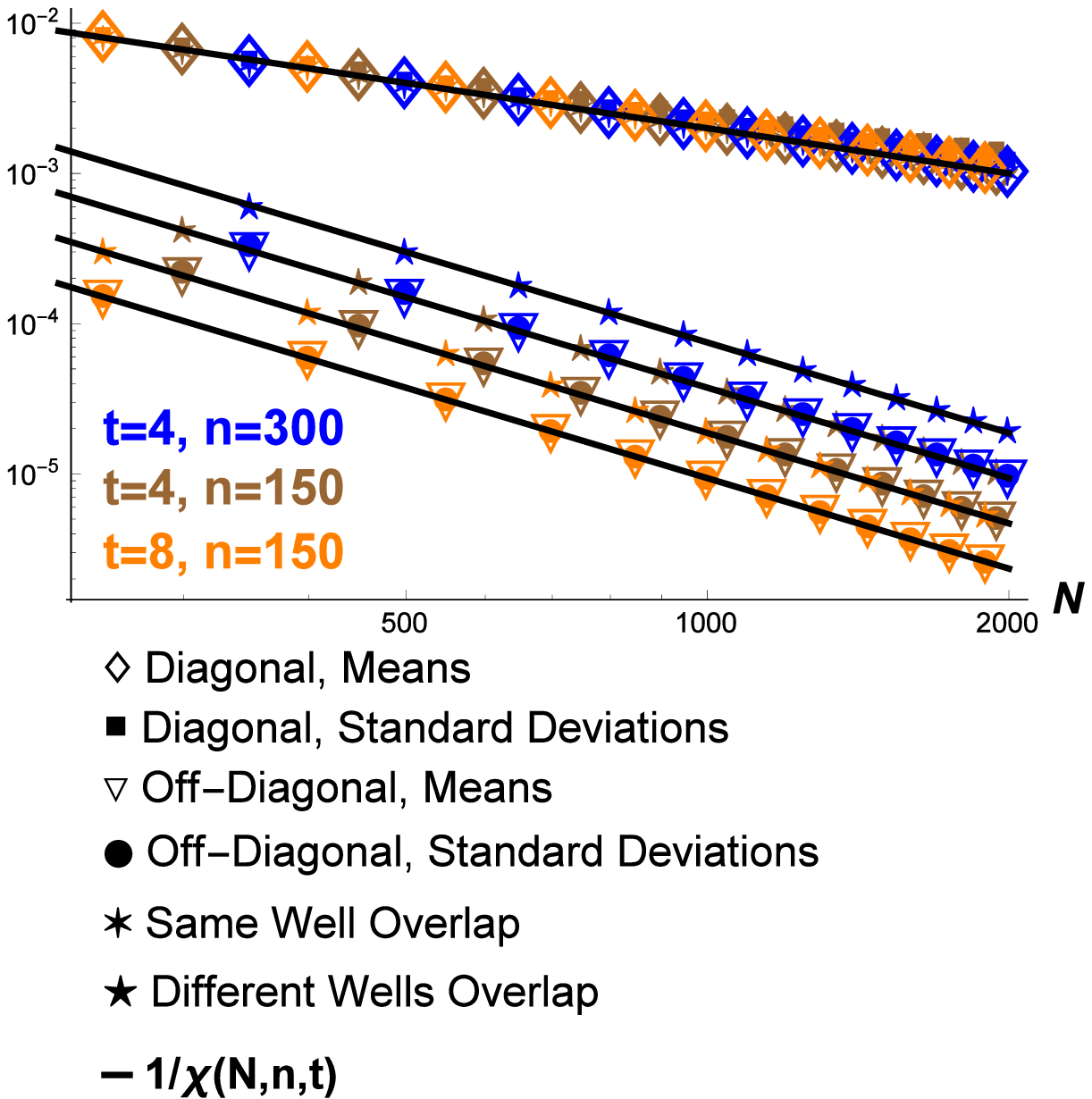}
\caption{Log-log plot of the means and standard deviations of the diagonal and off-diagonal blocks of the eigenvectors amplitudes matrices and of the means of the blocks of the overlap matrices. These results obtained by averaging over several realizations are in remarkable agreement with the analytical expectations (\ref{Dexp}, \ref{ODexp}), also plotted as continuous lines.}  
\label{fig:FitPlots}
\end{center}
\end{figure}

In Fig. \ref{fig:2BlocksU} we plot the squared absolute value of the entries a typical unitary matrix $\tilde {\bf U}$ constructed this way: we notice that it shows a characteristic block structure, with the off-diagonal blocks suppressed compared to the diagonal ones. This block structure indicates exactly that the eigenvectors have rotated mostly in the directions spanned by the eigenvectors corresponding to eigenvalues belonging to the same well. In Fig. \ref{fig:distrib} we plot the distributions of the diagonal and off-diagonal block entries of the matrix in fig. \ref{fig:2BlocksU}. We see that they both fit a Porther-Thomas distribution like (\ref{PTd}), but with 
\be
   \chi_{\rm D} = {N \over 2}  \quad {\rm and} \quad
   \chi_{\rm OD} =  {2 t N^2 \over n} 
  \label{mPTd}
\ee
for the diagonal and off-diagonal elements, respectively.
The former indicates that the eigenvectors are (almost perfectly) uniformly distributed over a $N/2$-dimensional sphere, while the latter is compatible with the expectation from a simple perturbation theory analysis.
Another important quantity to consider in the context of localization is the overlap between eigenstates, defined as $O_{j l}= \sum_m^N \left| \tilde U_{m j} \right|^2 \left| \tilde U_{m l} \right|^2$. In extended phases, the overlap scales like $1/N$: a faster decay indicates a suppression of the overlap and hence the onset of localization.
From the above matrix element distributions, we expect the overlap between eigenstates belonging to the same well (diagonal blocks of ${\bf O}$) and of different wells (off-diagonal blocks of ${\bf O}$) to be
\bea
   \langle \left| O_{jl} \right| \rangle_{\rm D} \: \: & = &
   \: \:  \langle \left| \tilde U_{jl} \right| \rangle_{\rm D} \; \;
   = \: \: \langle \left| \Delta \tilde U_{jl} \right| \rangle_{\rm D}  \; \; 
   = {1 \over \chi_{\rm D}} \; ,  
   \label{Dexp} \\
   \langle \left| O_{jl} \right| \rangle_{\rm OD} & = & 2 \langle \left| \tilde U_{jl} \right| \rangle_{\rm OD}  = 
   2 \langle \left| \Delta \tilde U_{jl} \right| \rangle_{\rm OD}  = {2 \over \chi_{\rm OD}} \; .
   \label{ODexp}
\eea
We checked these predictions against the numerical results obtained over several realizations of the applied perturbations in fig. \ref{fig:FitPlots}, concentrating in particular on the finite-size behavior, finding a remarkable agreement for all the quantities considered.

These results show that the off-diagonal blocks of the unitary matrix are suppressed by a power of $N$ compared to diagonal ones, indicating once more that in the planar limit (i.e., in the leading order in $N$) the full rotational symmetry is broken into a $U(N/2) \times U(N/2)$.

\section{Discussion \& Conclusions}

In this work, we have proven, both analytically and through a numerical experiment, that the rotational invariance of matrix models can be spontaneously broken, when the distribution of the eigenvalues has a disconnected support (the so-called  ``multi-cuts solutions'' \cite{demeterfi90, deift99, bleher05}). Although the mathematical literature has already discussed extensively the critical behavior of the eigenvalue distribution at the point where a gap in the distribution is about to open, our results characterizes it as a phase transition to a phase where the $U(N)$ symmetry has been broken to a smaller one. We provided a geometrical argument, an analytical symmetry breaking calculation (done in the spirit of \cite{moshe94}), and a numerical finite size analysis. These considerations, as well as the calculation shown in the {\it method} section, establishes once more the strict connection between the eigenvalue and eigenvector statistics. 

We discussed extensively the symmetric two-cuts case of a double well potential, but all our analysis can be extended straightforwardly to a potential resulting in any number of disjoint eigenvalue intervals. For a $n$-cut solution, in which each well contains $m_j$ eigenvalues, the $U(N)$ symmetry is broken into $\prod_{j=1}^n U(m_j)$ and each eigenvector becomes localized within a $m_j$-dimensional sphere, with spillage outside it suppressed and vanishing in the thermodynamic limit. The location of this sphere in the whole Hilbert space is random and uniform, similarly to the magnetization of a ferromagnet.  But once this direction is chosen, the system shows a certain rigidity against changing it, which gets completely frozen in the thermodynamic limit.

The problem of eigenvector localization in random matrices has already been discussed, by considering  non-invariant PDFs, (most noticeably, the so-called {\it Banded Random Matrices} \cite{mirlin96}), that is, by introducing an explicit basis dependence in the definition of the model. Unfortunately, these models are hardly treatable in an analytical way (with the exception of a  few perturbative approaches, valid only in certain regimes \cite{fyodorov91, fyodorov95, yevtushenko03, yevtushenko04}) and one has to rely mainly on numerics for the analysis. Our approach discloses the full power of the techniques available for invariant matrix models to the study of broken symmetries.

A lot of work needs now to be done: having established the existence of the SSB, the critical exponents of the (dis-)order parameter should be calculated. 
Most of all, a machinery to address eigenvector-related observables should be developed, to directly determine their localization properties. Such a machinery, for instance, can come from a replica approach, by measuring the overlaps between eigenvectors of different realizations of the same matrix. From this point of view, the spontaneous breaking of rotational symmetry translates into a replica symmetry breaking, with new, intriguing connections to be explored. 

We have established a direct connection between a certain SSB mechanism and localization in Hilbert space. The latter constitutes a long-standing theoretical problem, starting from the pioneering work by Anderson \cite{anderson58, khemani15} and arriving to the current activity in the fields of many-body localization (MBL) \cite{nandkishore15} or on network structure \cite{newman13}.  
Our work shows that localization is more than the mere vanishing of an eigenstates outside of a certain region: it is in fact a resistance of an eigenvectors from spreading under the effect of an external perturbation. This way of reasoning is consistent with the empirical fact that eigenvalue statistics (a base invariant quantity) is commonly used to discriminate a localized from a extended phase. Such base-independent characterization of ergodicity loss, based on the response to perturbations, can be especially useful to better understand Many-Body Localization. 

Another exciting prospect it to characterized the SSB of the so-called Weakly Confined Matrix Models (WCMM), introduced in \cite{muttalib93}, which have been shown to have the spectral properties of a system at the mobility edge of a metal/insulator transition (chapter 12 of  \cite{akemannbook}). The idea of a spontaneous breaking of $U(N)$ invariance originated from this observation \cite{canali95, pato00} and now our works paves the way to complete this analysis and to show that the eigenvectors of the WCMM become multifractal, rendering the WCMM the first analytically solvable toy-model capturing the universality of the Anderson transition.

Finally, the relevance of this SSB mechanism should be understood for matrix models emerging in field/string theories.
Matrix models arise frequently in string and M-theory settings, often as a result of a localization limit. Examples include several SUSY $U(N)$ gauge theories, topological models and variations such as ABJM (which is also related to the WCMM) \cite{natobook, akemannbook, polchinskibook2, marinobook}. These models are relevant both as fundamental theories, as well as for their holographic description of strongly interacting field theories in condensed matter systems or QCD. 
The SSB we discussed implies that a whole class of observables, which have been traditionally neglected because related to the matrix rotational degrees of freedom, can provide non-trivial information on these models, as it usually happens when a symmetry is found to be spontaneously broken.


The role of eigenvectors in matrix models has been long overlooked, but our results show that they have much to tell us.

\section{Acknowledgments}
I thank Prof. V. Kravtsov for introducing me to this class of problems and for the numerous discussions which guided me to these results. I also thank Prof. J. McGreevy for his inputs and support. I'm also in debt to Andrea Allais for his crucial help with the coding for the numerics.
This project has been supported by a Marie Curie International Outgoing Fellowship within the 7th European Community Framework Programme (FP7/2007-2013) under Grant PIOF-PHY-276093.

\appendix

\section{Methods}

\subsection{Methods: Symmetry Breaking Term}

To evaluate the effect of (\ref{genfunc}) we use the Itzykson-Zuber formula for the integration over the unitary group \cite{itzykson80}
\be
   \int \de {\bf U} \eu^{\tr {\bf A} {\bf U} {\bf B} {\bf U}^\dagger} 
   \propto {\det \left[ \eu^{a_j b_l} \right] \over \Delta \left( \{ a \} \right) \Delta \left( \{ b \} \right)} \; ,
  \label{IZformula}
\ee
where $\Delta \left( \{ a \} \right) = \prod_{j<l} (a_j - a_l)$ is the Van der Monde determinant constructed with the eigenvalues $\{a\}$ of ${\bf A}$. For its application to (\ref{genfunc}) we have ${\bf A} = {\bf M}$ and ${\bf B} = N J {\bf S}$.
The L.H.S. of eq. (\ref{IZformula}) is singular when two or more eingenvalues of either ${\bf A}$ or ${\bf B}$ coincide: both the determinant at the numerator and at the denominator vanish, because two or more rows (or columns) become identical. Since both singularities are of the same order, taking the limit where the eigenvalues become progressively close (point-splitting method), one achieves a finite expression (similar regularizations have been developed in \cite{kamenev99, fyodorov03}). If ${\bf B}$ has two sets of $N/2$ degenerate eigenvalues equal to $\pm b$ we have
\be
   \int \de {\bf U} \eu^{\tr {\bf A} {\bf U} {\bf B} {\bf U}^\dagger} \propto 
   {  1  \over \Delta \left( \{ a \} \right) } \: \det \left( \begin{array}{l} 
                  a_l^{j-1} \: \eu^{-b a_l}  \cr 
                  a_l^{j-1} \: \eu^{\: b a_l}  \cr 
                  \end{array} \right)_{l=1 \ldots N}^{j=1 \ldots N/2} \; .
   \label{IZ1} 
\ee
The determinant at the numerator can be further evaluated as
\be
   \det \left( \begin{array}{l} 
                  a_l^{j-1} \: \eu^{-b a_l}  \cr 
                  a_l^{j-1} \: \eu^{\: b a_l}  \cr 
                  \end{array} \right)
   = \sum_{\{ \alpha \} \cup \{ \alpha' \} = \{ a \} }^\prime
    \eu^{-b \sum_j \left( \alpha_j - \alpha_j^\prime \right)} \:
    \Delta \left( \{ \alpha \} \right) \Delta \left( \{ \alpha' \} \right) \; ,
   \label{IZ2}
\ee
where the sum is over all the partitions of the eigenvalues of ${\bf A}$ into two sets of $N/2$ elements, such that each partition can be obtained from an even permutation of the original elements.
We see that the degeneracies of the eigenvalues of ${\bf B}$ induces a suppression of the interaction between the {\bf A} eigenvalues, in the form of a reduced Van der Monde determinant.

Combining these formulas into (\ref{genfunc}), we see that the effect of the symmetry breaking term is to isolate one such partition (corresponding to ${\bf \Lambda} {\bf T}$), which amounts to assigning each eigenvalue to a given well, and to suppress the interaction between eigenvalues of different wells. This interaction is progressively restored by the including the other partitions in (\ref{IZ2}): each of them represent a configurations in which pairs of eigenvalues have switched wells. However, these configurations contribute to the partition function with a weight given by $\exp \left( -2 b \left| \alpha_j - \alpha'_j \right| \right)$ for each pair of eigenvalues switched, compared to the leading configuration. These contributions are suppressed for large $b = N J \tau$ or large eigenvalue distance and disappear in the thermodynamic limit.

\subsection{Methods: Effect of a perturbation}

In the numerical experiment, we solve the plasma equation for the eigenvalues, to determine their equilibrium distribution under the effect of the potential (\ref{V2W}).
Having determined ${\bf \Lambda}$, we obtain a typical matrix ${\bf M} = {\bf U}^\dagger {\bf \Lambda} {\bf U}$ of the double well matrix ensemble by generating a unitary matrix ${\bf U}$ with the algorithm in \cite{mezzadri06}.
As a perturbation, we generate ${\bf \Delta M}$, a sparse, Hermitian random Gaussian matrix, with $n \times N$ non-zero elements. We determine the unitary matrix ${\bf U'}$ which diagonalizes ${\bf M} + {\bf \Delta M}$ and study it in the basis where ${\bf M}$ is diagonal: that is, we are looking at $\tilde{\bf U} = {\bf U}' {\bf U}^\dagger$.
For Fig. \ref{fig:FitPlots}, we generated the equilibrium distribution of eigenvalues for $t=4, 8$ and consider perturbations with $n=150, 300$. Furthermore, each data point is the result of averaging over several realizations (we took $30$ realizations for $N< 700$, $20$ for $N<1200$, and $10$ otherwise, noticing that each realization is already quite accurately self-averaging).

\end{document}